# A Field-Deployable 2U Ultra-Stable Laser with Sub-Hertz Linewidth Based on Optoelectronic Oscillator Locking

Dongdong Jiao[1,3,4], Linbo Zhang[1], Xue Deng[5], Jing Gao[1,4], Yue Gao[1,4], Mengfan Wu[1], Qi Zang[1], Guanjun Xu[1,2,3,4,**], and Tao Liu[1,2,3,4,**]

[1] National Time Service Center, Chinese Academy of Sciences, Xi'an 710600, China

[2] Key Laboratory of Time Reference and Application, Chinese Academy of Sciences, Xi'an 710600, China

[3] Hefei National Laboratory, Hefei 230088, China

[4] University of Chinese Academy of Sciences, Beijing 100049, China

[5] Communications and Information Engineering, Xi'an University of Posts and Telecommunications, Xi'an 710121, China

**Corresponding authors. Email: xuguanjun@ntsc.ac.cn; taoliu@ntsc.ac.cn

**Abstract:** Compact, high-performance ultra-stable lasers that can operate outside the laboratory remain a long-standing goal. Here we report, to our knowledge, the most compact field-deployable ultra-stable laser, packaged in a standard 2U rack chassis. It is stabilized to a 50 mm-long, 19 mL optical reference cavity with low vibration sensitivity by optoelectronic oscillator (OEO) locking, an approach particularly suited to compact, transportable systems. The laser delivers an integrated linewidth of ~0.42 Hz and a fractional frequency stability of ~$5\times10^{-15}$ at 1 s averaging time (~$2.2\times10^{-15}$ at 0.4 s). Measured phase noise approaches the cavity thermal-noise limit below 100 Hz and the theoretical OEO-locking limit from 100 Hz to 100 kHz. The system volume is roughly one quarter of the volume of previously reported integrated, transportable ultra-stable lasers of comparable performance, making it well suited for field applications in time-frequency metrology, quantum technologies, and environmental sensing.

## 1. Introduction

Ultra-stable lasers underpin numerous advanced applications, including optical clocks [1,2], ultra-stable microwave generation [3,4], gravitational wave detection [5,6], environmental sensing [7,8], dark matter detection [9], quantum communication [10,11], fiber time-frequency transfer [12,13], geodesy [14], free-space optical links [15], seismic monitoring [16], and photonics-based radar [17,18]. Among existing laser frequency stabilization techniques, Pound-Drever-Hall (PDH) locking is the most widely used to stabilize lasers to optical reference cavities [19], owing to its robustness and high signal-to-noise ratio [20,21]. State-of-the-art PDH-stabilized lasers have achieved fractional frequency stability at the $10^{-17}$ level and millihertz-level linewidths [22-24].

Although the PDH locking scheme offers outstanding performance, it has intrinsic limitations. As a frequency-discriminating technique, its error signal responds to the laser frequency deviation and imposes stringent requirements on the seed laser linewidth [19-21]. Moreover, the limited high-frequency loop gain of PDH servo systems tends to introduce servo bumps that degrade the high-frequency phase noise [25,26]. These constraints have hindered the deployment of cavity-stabilized lasers in applications that urgently require miniaturized and field-deployable instruments, such as space optical clocks [27], space-borne gravitational wave detection [28], and field geodesy [14].

Laser locking based on an optoelectronic oscillator (OEO) offers an attractive alternative for addressing these bottlenecks. First demonstrated in 1998 [29] and 2000 [30], the technique has attracted renewed interest [25,26], fueled by new measurement methods, optical feedforward noise cancellation, and refined theoretical modeling. In OEO locking, the laser phase noise is transferred onto the phase of the radio-frequency carrier inside the optoelectronic oscillation loop. The scheme operates as a phase-discriminating lock, in which the error signal responds to phase fluctuations, yielding superior low-frequency noise suppression compared with PDH [25,26]. Experiments have shown that OEO locking can suppress the noise of a DFB laser by more than 140 dB at a 10 Hz offset, about 70 dB better than PDH under identical hardware conditions [25]. In particular, OEO locking can directly lock broad-linewidth DFB lasers down to the thermal noise limit of the optical reference cavity, which suits miniaturized laser systems; Feedforward correction can further compensate for the noise left by the limited loop bandwidth, suppressing servo bumps and improving the high-frequency phase noise [25]. Recent work has confirmed that OEO-stabilized lasers can approach the cavity thermal noise limit with a fractional frequency stability at the $10^{-15}$ level [26].

However, reported experimental implementations of OEO laser locking remain laboratory tabletop systems [25,26], with neither an integrated instrument suited to field deployment nor an automatic frequency-locking function. Here we develop a field-deployable ultra-stable laser based on OEO locking and integrate it into a standard 2U rack chassis. A compact cylindrical reference cavity of 50 mm length and 19 mL volume is designed for low vibration sensitivity, and its passive thermal shields and compact vacuum subsystem are optimized to suppress ambient temperature disturbances while minimizing size. A microcontroller-based hybrid analog-digital controller realizes fully automatic frequency locking, enabling unattended operation. The laser is characterized by optical heterodyne beating against a second ultra-stable laser with a fractional frequency stability of $10^{-16}$-level at 1 s, yielding a beat note linewidth of ~0.95 Hz and an integrated linewidth of ~0.42 Hz. Phase-noise measurements show that the stabilized laser approaches the thermal-noise limit of the reference cavity below 100 Hz and follows the theoretical OEO-locking limit from 100 Hz to 100 kHz. The system achieves a fractional frequency stability of $\sim 5\times10^{-15}$ at 1 s averaging time ($\sim 2.2\times10^{-15}$ at 0.4 s). To our knowledge, this is the most compact field-deployable ultra-stable laser to date, with a total volume approximately one quarter that of previously reported integrated, transportable ultra-stable lasers of comparable performance.

This paper is organized as follows. Section 2 describes the fundamental principle and experimental setup. Section 3 details the system design, including the vacuum assembly, optical reference cavity, and automatic locking system. Section 4 presents experimental characterization of phase noise, linewidth and frequency stability. Section 5 concludes this work.

## 2. Fundamental principle and experimental setup

The frequency-stabilization principle of the proposed ultra-stable laser is illustrated in Fig. 1. A commercial fiber laser operating at 1550.12 nm with a free-running linewidth of ~1 kHz is stabilized to an optical reference cavity with a finesse of ~316,000 using the optoelectronic oscillator (OEO) locking scheme proposed and analyzed in Refs. [25,26]. The laser output (~20 mW) is frequency-shifted by +50 MHz with a polarization-maintaining (PM) fiber-coupled acousto-optic modulator (AOM) that serves as the fast frequency actuator. The first-order diffracted beam from the AOM is split by a 90/10 PM fiber coupler: 90% of the optical power provides the ultra-stable output for characterization and future applications, and the remaining 10% feeds the locking path. There, the beam is phase-modulated by a fiber electro-optic modulator (EOM) with a half-wave voltage of 4 V and a bandwidth of 10 GHz. The carrier and modulation sidebands then pass through a PM fiber circulator, and ~50 μW of light is coupled into the cavity. Since only the lower sideband is cavity-resonant, the light reflected from the cavity yields, upon detection by a photodetector (PD), a ~158 MHz beat note between the carrier and this sideband; the photocurrent is amplified by a low-noise amplifier (LNA) with 40 dB gain. The resulting radio-frequency (RF) signal is divided by a power splitter into two branches: one supplies the RF drive of the EOM, closing the OEO loop, and the other passes through a divide-by-10 frequency divider and is mixed with an external reference RF source to generate the error signal. This error signal feeds two servo controllers: Servo 1 tunes the laser piezoelectric transducer (PZT) voltage, and Servo 2 adjusts the AOM driving frequency. To suppress residual amplitude modulation, PM components and PM fusion splicing are used throughout the fiber loop, achieving a degree of polarization (DOP) above 95% for the entire optical path.

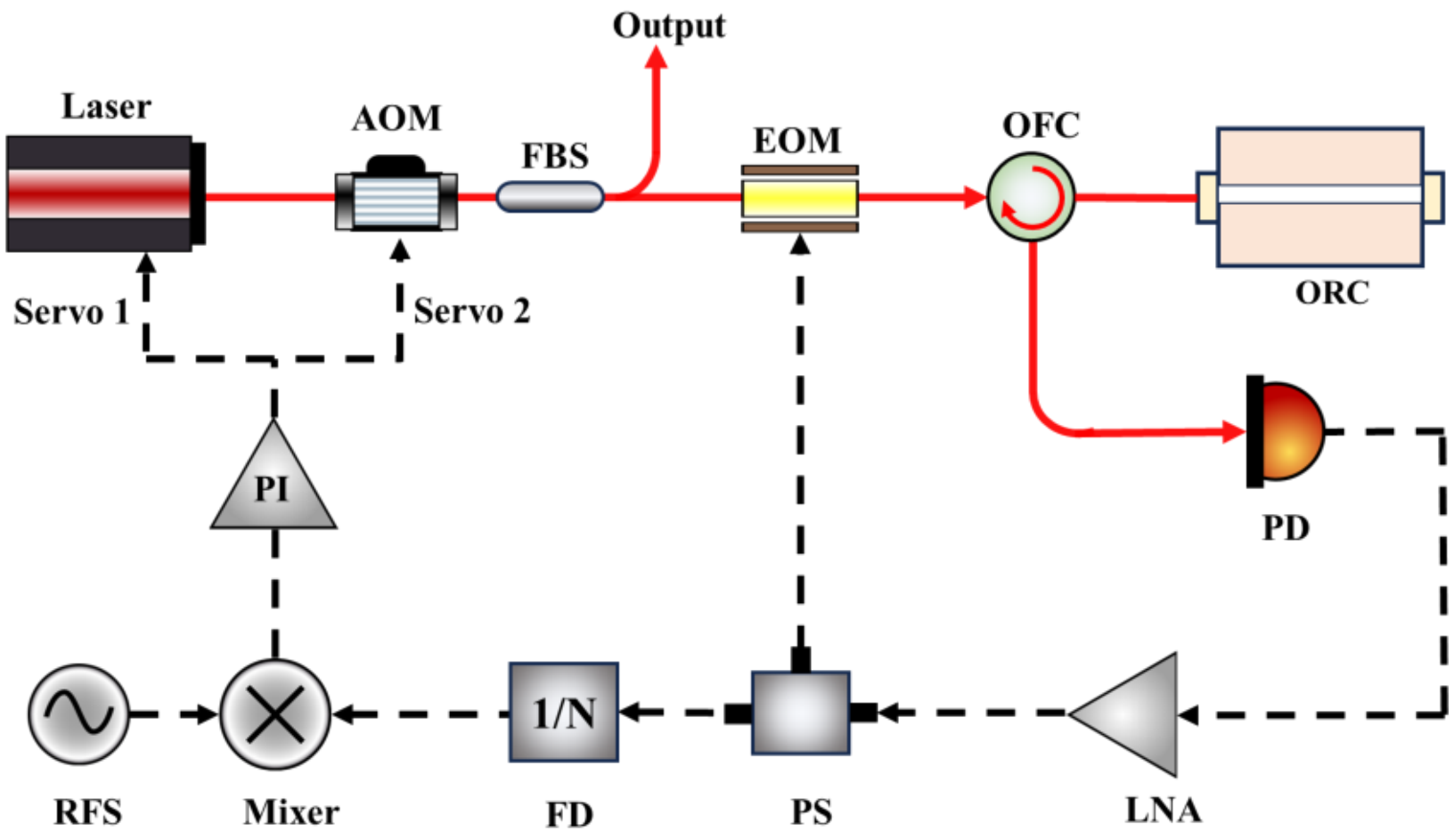

Fig.1 The schematic diagram showing the basic principle of an ultra-stable laser. Laser: fiber-based narrow-linewidth laser, AOM: acousto-optic modulator, FBS: fiber beam splitter, EOM: electro-optic modulator, OFC: optical fiber circulator, ORC: optical reference cavity, PD: photodetector, LNA: low-noise amplifier, PS: power splitter, FD: frequency divider, RFS: radio-frequency source, PI: servo controller.

The theoretical framework for OEO-locked ultra-stable lasers has been reported by Q. Cen *et al.* [26] and T. Nakamura *et al.* [25]. Herein, we present an optical-field-based theoretical model to describe this locking mechanism. The optical field of a free-running continuous-wave laser is written as:

$$E_0(t) = E_0 e^{i(\omega_{Las} t + \varphi_0)} \tag{1}$$

where $E_0$ is the electric field amplitude, $\omega_{Las}$ denotes the laser central angular frequency, and $\varphi_0$ represents the time-dependent random laser phase. The laser illuminates the EOM, which is driven by a radio-frequency signal regenerated from the OEO loop at a modulation angular frequency $\Omega_{OEO}$ with a modulation index $\beta$. Under the small-signal-modulation approximation, only the 0th-order and $\pm$1st-order Bessel sidebands are retained. The optical field after the EOM becomes:

$$E_{EOM}(t) = E_0 J_0(\beta) e^{i(\omega_{Las} t + \varphi_0)} + E_0 J_1(\beta) e^{i(\omega_{Las} t + \Omega_{OEO} t + \varphi_0 + \varphi_{OEO})} - E_0 J_1(\beta) e^{i(\omega_{Las} t - \Omega_{OEO} t + \varphi_0 - \varphi_{OEO})} \tag{2}$$

where $J_0(\beta)$ and $J_1(\beta)$ are the 0th-order and 1st-order Bessel functions of the first kind, respectively, and $\varphi_{OEO}$ is the phase of the OEO radio-frequency drive signal. Let $F(\omega)$ represent the reflection function of the optical reference cavity. The optical field incident on the PD is then:

$$E_{ref} = F(\omega_{Las})E_0J_0(\beta)e^{i(\omega_{Las}t+\varphi_0)} + F(\omega_{Las}+\Omega_{OEO})E_0J_1(\beta)e^{i(\omega_{Las}t+\Omega_{OEO}t+\varphi_0+\varphi_{OEO})} - F(\omega_{Las}-\Omega_{OEO})E_0J_1(\beta)e^{i(\omega_{Las}t-\Omega_{OEO}t+\varphi_0-\varphi_{OEO})} \quad (3)$$

Only the lower sideband satisfies cavity resonance, whereas the carrier and upper sideband are strongly detuned; accordingly, $F(\omega_{Las})$ and $F(\omega_{Las}+\Omega_{OEO})$ can be approximated as unity. The PD photocurrent is expressed as:

$$\begin{cases} i(t) = P_C + P_S(1+|F(\omega_{Las}-\Omega_{OEO})|^2) + \sqrt{P_CP_S}\big(1-F^*(\omega_{Las}-\Omega_{OEO})\big)e^{i(\Omega_{OEO}t+\varphi_{OEO})} \\ \qquad +O(2\Omega_{RF}) \\ P_C = |E_0J_0(\beta)|^2 \\ P_S = |E_0J_1(\beta)|^2 \\ \sqrt{P_CP_S} = |E_0|^2J_0(\beta)J_1(\beta) \end{cases}$$

(4)

Here $P_C$ and $P_S$ denote carrier power and sideband power, respectively, and $F^*(\omega_{Las}-\Omega_{OEO})$ is the complex conjugate of $F(\omega_{Las}-\Omega_{OEO})$. We incorporate OEO loop phase noise $\varphi_{OEO-loop}$ (originating from photocurrent shot noise and electrical amplification noise) together with the phase delay $\Omega_{OEO}\tau_{OEO}$ induced by loop propagation delay $\tau_{OEO}$. The radio-frequency photocurrent component oscillating at $\Omega_{OEO}$ is:

$$i_{RF}(t) = \sqrt{P_CP_S}\big(1-F^*(\omega_{Las}-\Omega_{OEO})\big)e^{i(\Omega_{OEO}t+\varphi_{OEO}+\varphi_{OEO-loop}-\Omega_{OEO}\tau_{OEO})} \quad (5)$$

After amplification, this radio-frequency signal drives the EOM to close the regenerative feedback loop. At steady-state oscillation, the total phase accumulated over one full loop round-trip reproduces the initial phase condition, yielding the OEO angular oscillation frequency:

$$\Omega_{OEO} = \frac{\tau_{cav}}{\tau_{cav}+\tau_{OEO}}(\omega_{Las}-\omega_{cav}) + \frac{\varphi_{OEO-loop}}{\tau_{cav}+\tau_{OEO}} \quad (6)$$

In this expression, $\tau_{cav}$ is the photon decay time inside the optical reference cavity and $\omega_{cav}$ is the cavity resonant angular frequency. Assuming uncorrelated fluctuations for $\omega_{Las}$, $\omega_{cav}$, and $\varphi_{OEO-loop}$, we derive the OEO locking phase-noise power spectral density $S_{OEO}^{\phi}(f)$:

$$S_{OEO}^{\phi}(f) = \left(\frac{\tau_{cav}}{\tau_{cav}+\tau_{OEO}}\right)^2\left(S_{Las}^{\phi}(f)+S_{cav}^{\phi}(f)\right) + \frac{S_{\varphi_{OEO-loop}}^{\phi}(f)}{\omega^2(\tau_{cav}+\tau_{OEO})^2} \quad (7)$$

Under the condition $\tau_{cav} \gg \tau_{OEO}$, the phase-noise spectral density for the feedforward-corrected output becomes:

$$S_{OEO}^{\phi-F}(f) \approx \left(\frac{\tau_{OEO}}{\tau_{cav}}\right)^2 S_{Las}^{\phi}(f) + S_{cav}^{\phi}(f) + \frac{S_{\varphi_{OEO-loop}}^{\phi}(f)}{(\omega\tau_{cav})^2} \quad (8)$$

Equation (8) shows that the intrinsic laser phase noise is suppressed by a factor of $(\tau_{OEO}/\tau_{cav})^2$, so that the system output noise approaches the thermal noise limit of the optical reference cavity. For our hardware implementation, the estimated loop delay $\tau_{OEO}$ is approximately 10 ns.

## 3. Design of the ultra-stable laser system

The overall system comprises four major subsystems: a high-vacuum physics assembly, an optical reference cavity, an optical unit containing the pre-cavity and post-cavity optics, and an electronics unit.

### 3.1. Vacuum physics assembly

The vacuum physics assembly provides a stable, well-controlled environment for the optical reference cavity, suppressing temperature variations and perturbations from residual gas. As shown in Fig. 2, it consists of a vacuum chamber, the optical reference cavity, an ion pump, a vacuum feedthrough pin, a thermal shield, and two fiber collimators. Its outer envelope measures approximately 140 mm × 80 mm × 40 mm. A 0.5 L/s ion pump maintains the chamber pressure at the $10^{-5}$ Pa level. The feedthrough pin delivers power and electrical signals to the thermistor and heating film used for active temperature regulation inside the chamber; these components are mounted on the outer surface of the thermal shield and secured with polyimide tape. A gold-plated aluminum-alloy passive thermal shield surrounds the cavity. The cavity assembly is fastened by screws to four support posts rigidly attached to the chamber base, and PEEK pads are inserted between the screws and the cavity to provide thermal isolation. Two fiber collimators are mounted on the optical axis at the two ends of the chamber for laser input and output.

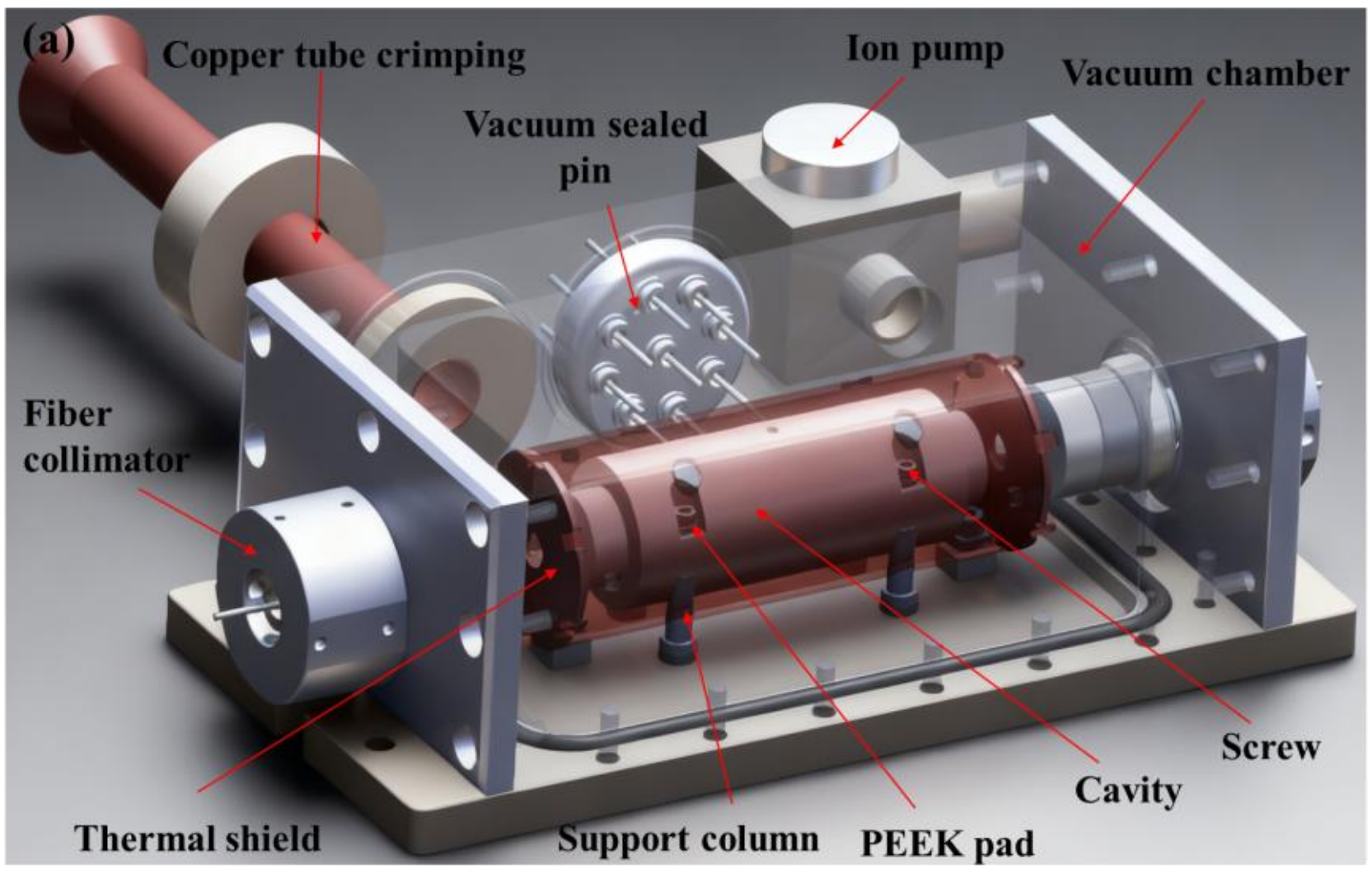

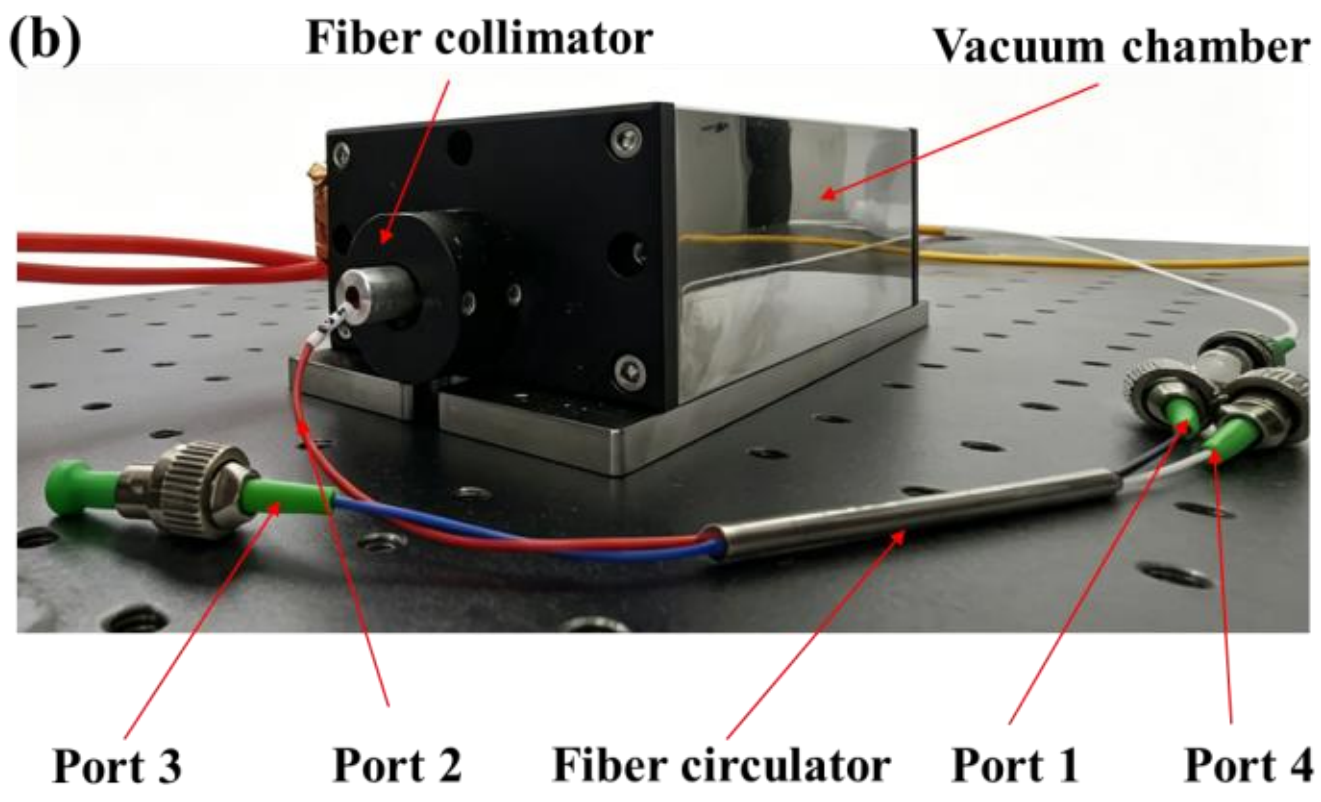

Fig.2 Schematic diagram and photograph of the vacuum physics assembly containing the optical reference cavity. (a) Perspective view. (b) Photograph.

## 3.2. Optical reference cavity

### Fundamental design

The optical reference cavity investigated in this work is illustrated in Fig. 3. It consists of two high-reflectivity mirrors optically bonded to a cylindrical spacer. The spacer has a length of 50 mm and a diameter of 20 mm. Both the mirror substrates and the spacer are fabricated from ultra-low-expansion (ULE) glass; the key material parameters are summarized in Table 1. The mirrors carry ion-beam-sputtered $SiO_2/Ta_2O_5$ dielectric multilayers optimized for 1550 nm, with a total coating thickness of ~7 μm. The mirror has a radius of curvature of 1 m; the additional geometric dimensions are listed in Table 2. Four symmetric mounting shoulders with fastening holes are machined on the spacer for mechanical attachment. A central bore of 5 mm diameter runs through the spacer for light propagation, and a vent hole located midway along the spacer allows the cavity interior to be evacuated.

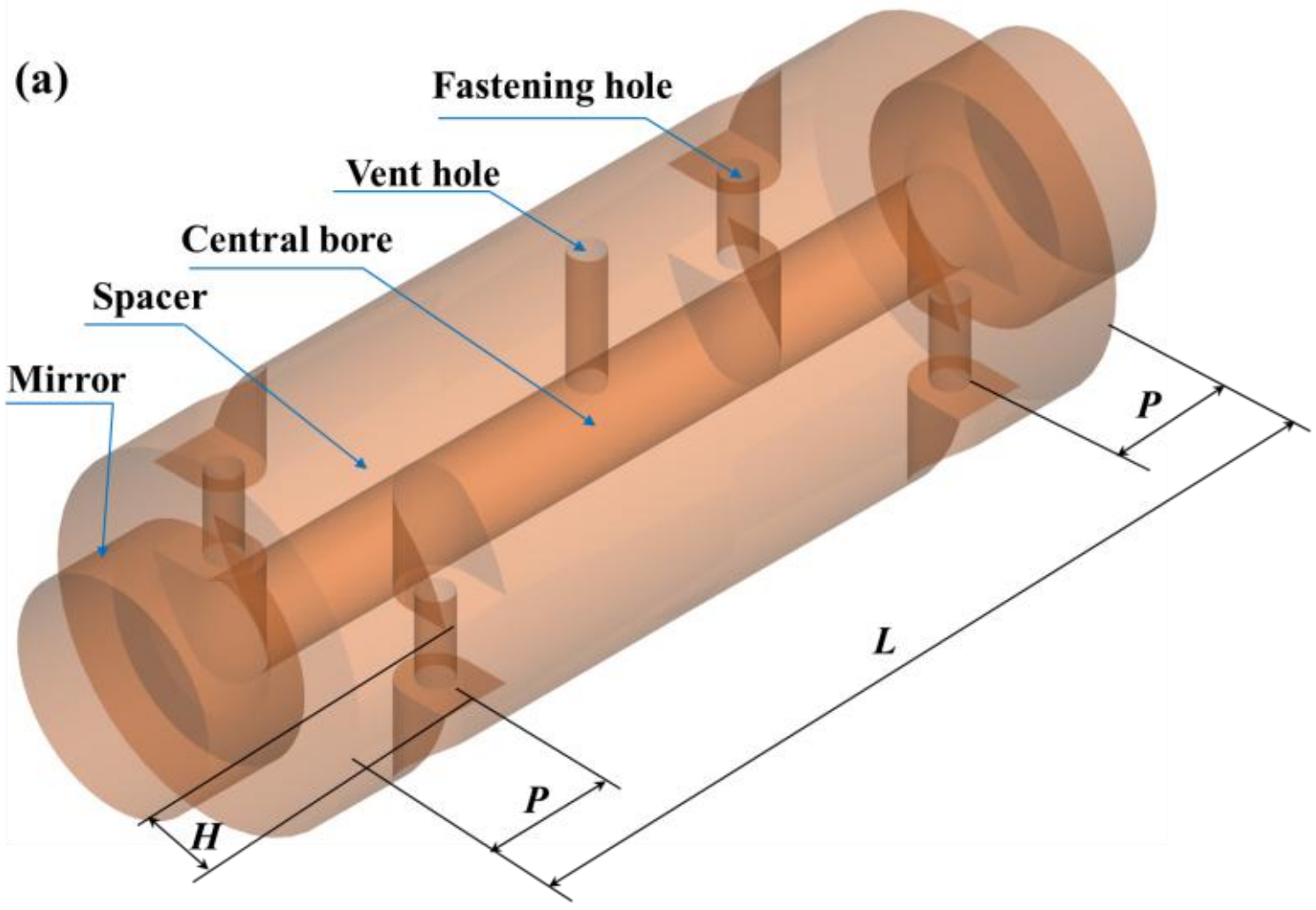

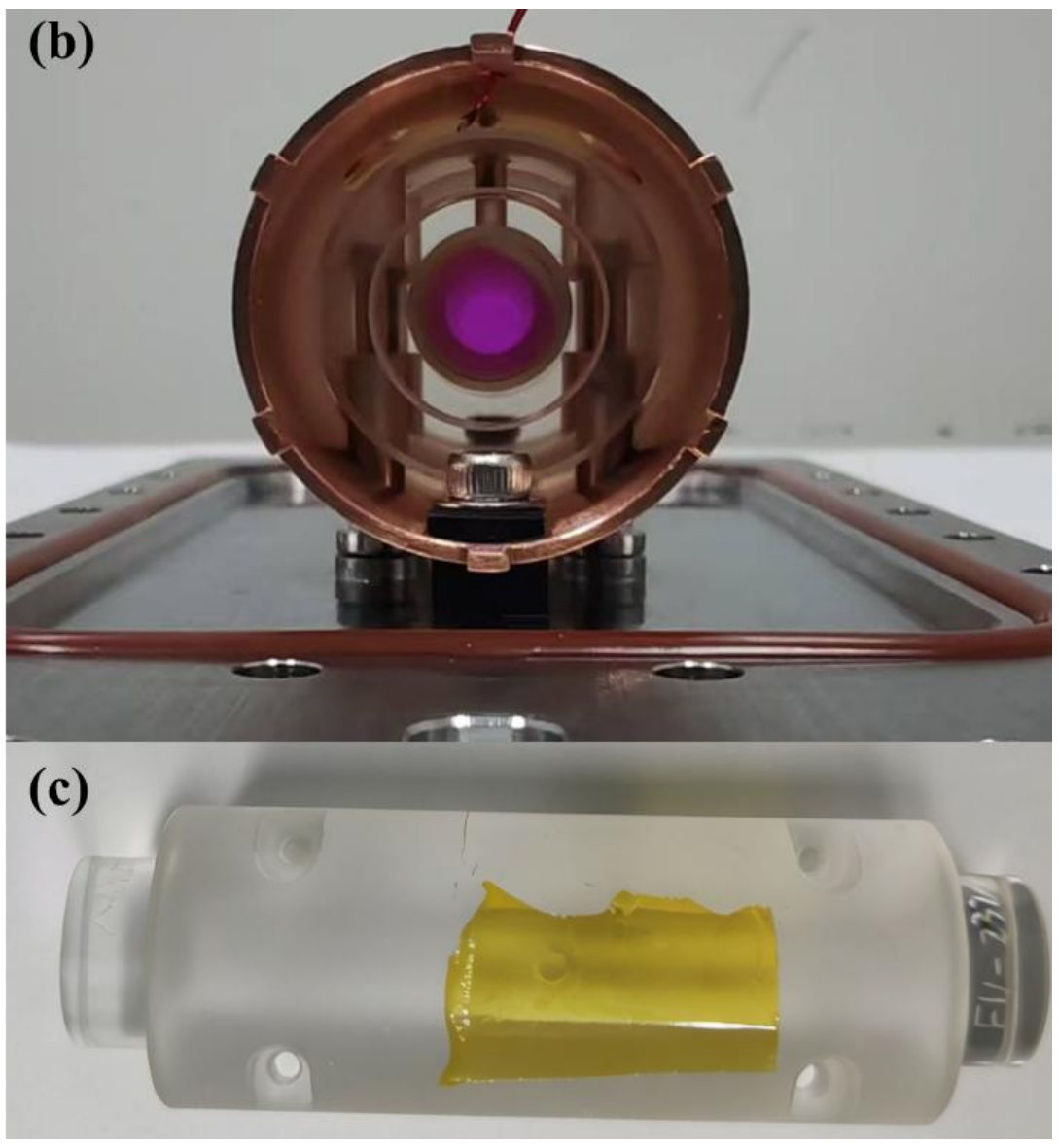

Fig.3 Optical reference cavity. (a) Three-dimensional diagram. (b) Photograph of the cavity assembly. (c) Photograph of the optical reference cavity.

Table 1 Mechanical performance parameters of ULE materials

| Type | Density (kg/m$^3$) | Young's modulus (GPa) | Poisson ratio | Loss angle |
|---|---|---|---|---|
| Symbol | $\rho$ | $E$ | $\sigma$ | $\phi$ |
| Value | 2210 | 67.9 | 0.17 | $1.7\times10^{-5}$ |

Table 2 Geometric parameters of the optical reference cavity

| Index | Symbol | Type | Value |
|---|---|---|---|
| 1 | $L$ | Cavity length | 50 mm |
| 2 | $D$ | Spacer diameter | 20 mm |
| 3 | $D_m$ | Mirror diameter | 12.7 mm |
| 4 | $h_m$ | Mirror thickness | 6.3 mm |
| 5 | $d_c$ | Coating thickness | 7 μm |
| 6 | $D_s$ | Spacer central bore diameter | 5 mm |
| 7 | $w$ | Beam waist | 250 μm |
| 8 | $D_v$ | Vent hole diameter | 2 mm |
| 9 | $D_f$ | Fastening hole diameter | 2 mm |
| 10 | $H$ | Fastening hole thickness | 5 mm |

## Finesse and zero-expansion temperature

Cavity finesse is a critical figure of merit for ultra-stable laser systems: according to Eqs. (7) and (8), the finesse strongly influences the OEO locking accuracy and the noise-rejection capability. The finesse depends only on the mirror reflectivity and is independent of the cavity length, incident optical power, and laser frequency. Common approaches to measuring the finesse include cavity ring-down, frequency sweeping, and linewidth scanning, of which linewidth scanning offers the highest precision. Our cavity was characterized by the linewidth-scanning method. As shown in Fig. 4, a Lorentz fit yields a cavity linewidth of ~9.5 kHz, corresponding to a finesse of ~316,000. The measured optical coupling efficiency into the cavity is approximately 40%.

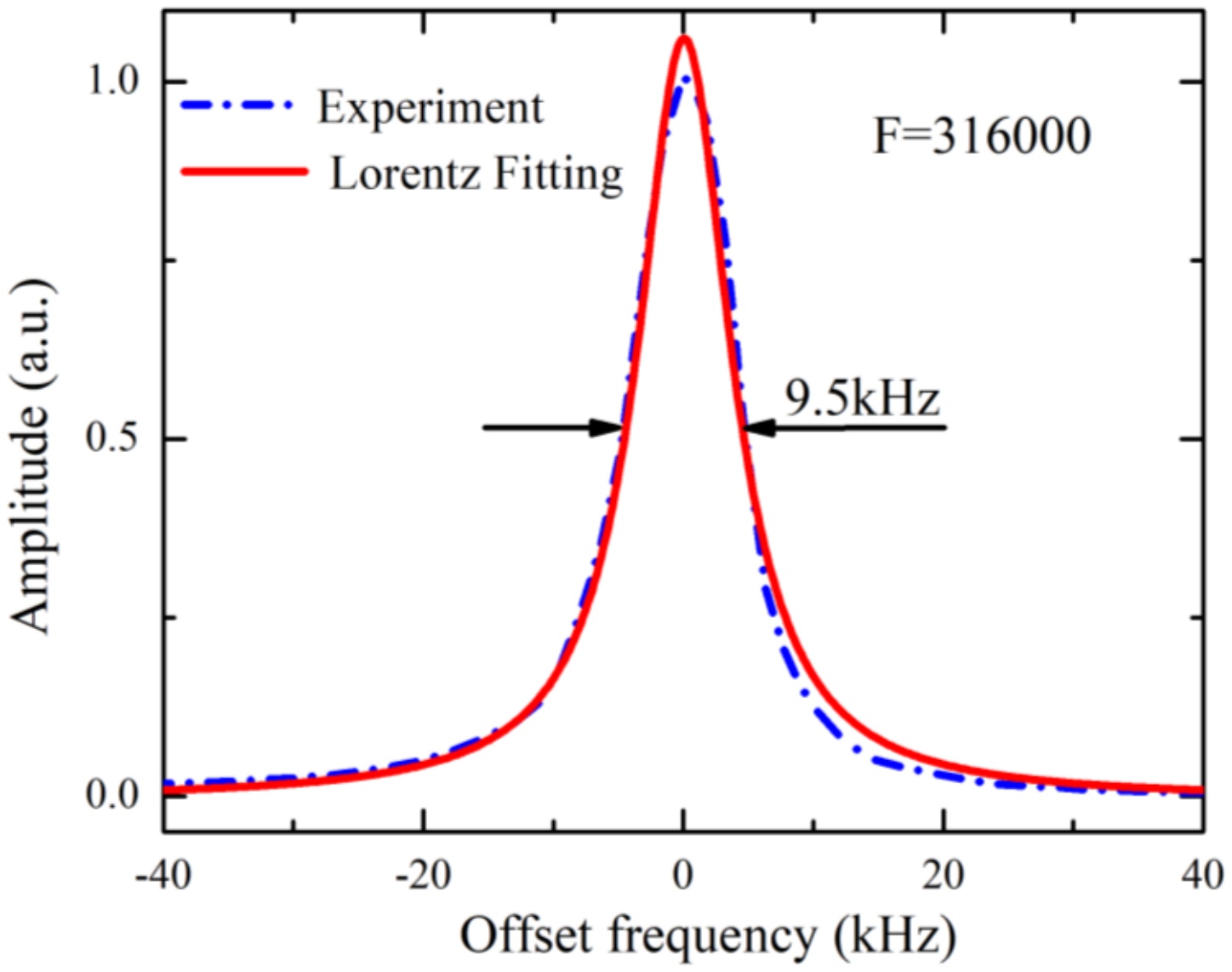

Fig.4 Measured cavity linewidth and finesse of the optical reference cavity. Blue dash-dotted line: experimental data; red solid line: Lorentz fit.

Temperature perturbations change the physical cavity length and thereby degrade the laser frequency stability. Two complementary strategies were adopted to mitigate thermal effects: operating the cavity near the zero-crossing temperature (ZCT) of its coefficient of thermal expansion, and suppressing the amplitude of the temperature fluctuations experienced by the cavity. The ZCT was determined by measuring the cavity resonant frequency as a function of temperature and locating the turning point of a quadratic fit. Following the procedure described in Ref. [31], we recorded the cavity response at six discrete temperatures and performed a second-order polynomial fit, which yields a ZCT of approximately 17.5 °C for this ULE cavity, as shown in Fig. 5. Between 15 °C and 20 °C, the cavity thermal expansion coefficient remains on the $10^{-9}$/K (ppb/K) level

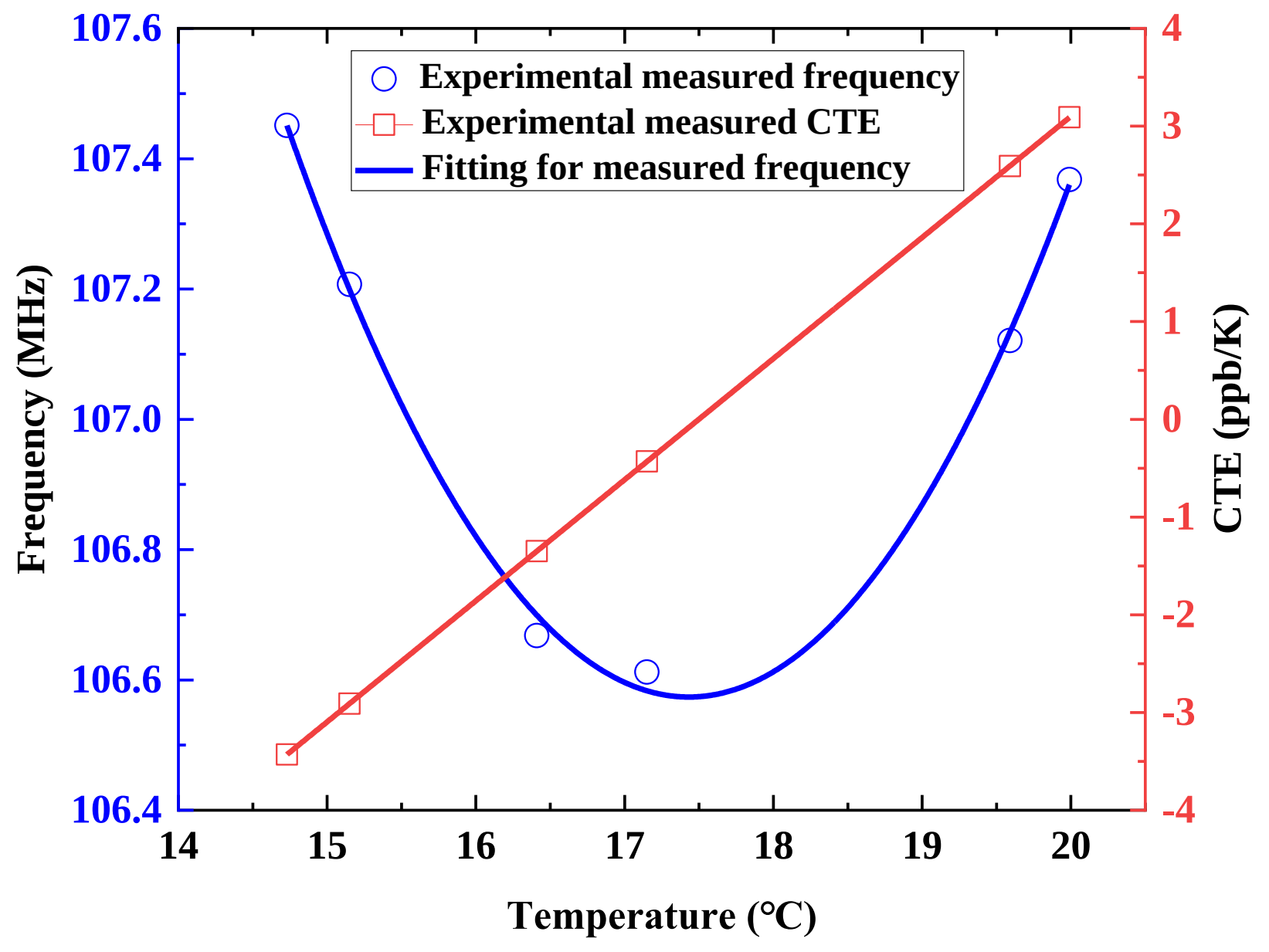

Fig.5 Measured linear thermal expansion coefficient of the optical reference cavity.

Assuming an offset of 0.1 °C between the actual set point and the ideal zero-crossing temperature, the temperature fluctuation experienced by the cavity must be confined to the 1–10 μK range to keep the temperature-induced fractional frequency instability better than $1\times10^{-15}$. Direct stabilization of the cavity body to microkelvin-level fluctuations is technically demanding, so multiple thermal shields surround the cavity, with active temperature control applied to the shields to isolate the cavity from ambient variations. The cavity is housed in a sealed vacuum chamber equipped with active temperature control. The thermal low-pass filtering behaviour of the vacuum assembly is described by [32]:

$$\frac{\Delta T_{cavity}(f)}{\Delta T_{out}(f)} \approx \frac{1}{f\tau} \tag{9}$$

Here $\Delta T_{out}(f)$ is the external temperature fluctuation, $\Delta T_{cavity}(f)$ is the fluctuation experienced by the cavity inside vacuum, $f$ is the Fourier frequency, and $\tau$ is the thermal time constant from the vacuum chamber to the cavity.

Equation (9) indicates that a larger thermal time constant suppresses cavity temperature fluctuations more effectively. Accordingly, we implemented a passive thermal shield of aluminum alloy, ~1 mm thick and gold-plated on both its inner and outer surfaces to reduce the emissivity. A finite-element thermal simulation (Fig. 6) predicts a thermal time constant of ~40,000 s, so that the residual thermal contribution to the 1 s fractional frequency instability is expected to be at the $10^{-16}$ level.

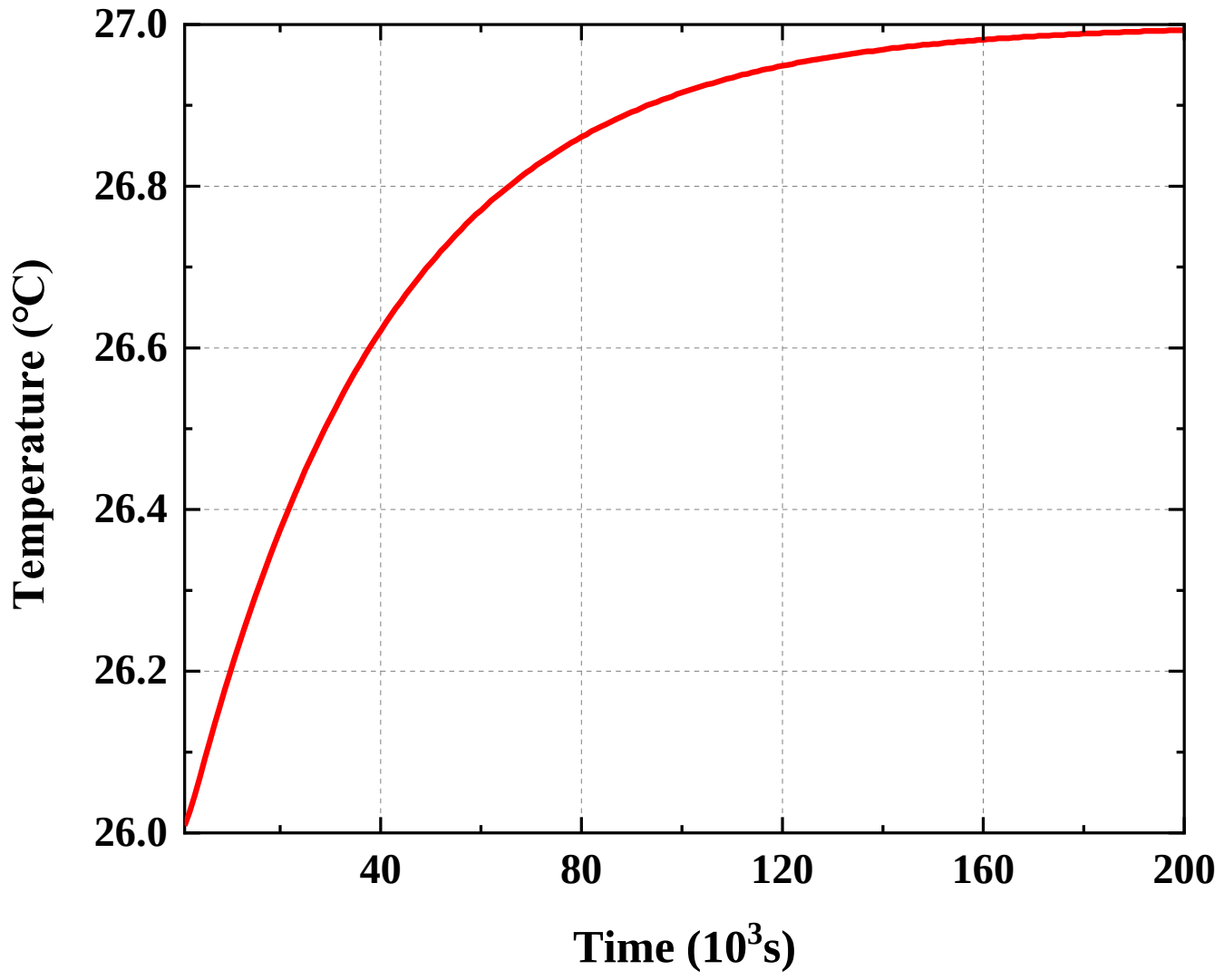

Fig.6 Simulated thermal time constant $\tau$ from the vacuum chamber to the optical reference cavity.

## Vibration sensitivity optimization

Vibration induces stress inside the optical reference cavity, modifying the cavity length and producing frequency shifts that degrade the laser stability. Vibration-induced degradation can be mitigated either by attenuating the incoming vibration noise or by minimizing the intrinsic vibration sensitivity of the cavity assembly. The vibration sensitivity $\alpha_v$ is determined by the cavity geometry and the mounting boundary conditions, and can be optimized by finite-element simulation [33,34]. It is defined as:

$$\begin{cases} \Delta L = L_{F+2g} - L_{F+1g} \\ \alpha_v = \frac{\Delta L}{Lg} \end{cases} \tag{10}$$

where $L$ is the nominal cavity length, $\Delta L$ is the length change, g is the gravitational acceleration, $L_{F+1g}$ is the cavity length under preload force $F$ plus 1g acceleration, and $L_{F+2g}$ corresponds to preload $F$ plus 2g acceleration.

Key parameters governing vibration sensitivity include distance $P$ from mounting holes to cavity end faces, mounting shoulder thickness $H$, and assembly clamping force $F$. ULE glass is brittle; hence clamping force is constrained between 1 N and 50 N, and mounting-hole thickness $H$ is fixed at 5 mm. We perform finite element analysis to optimize the critical parameter $P$. Simulation results are plotted in Fig. 7.

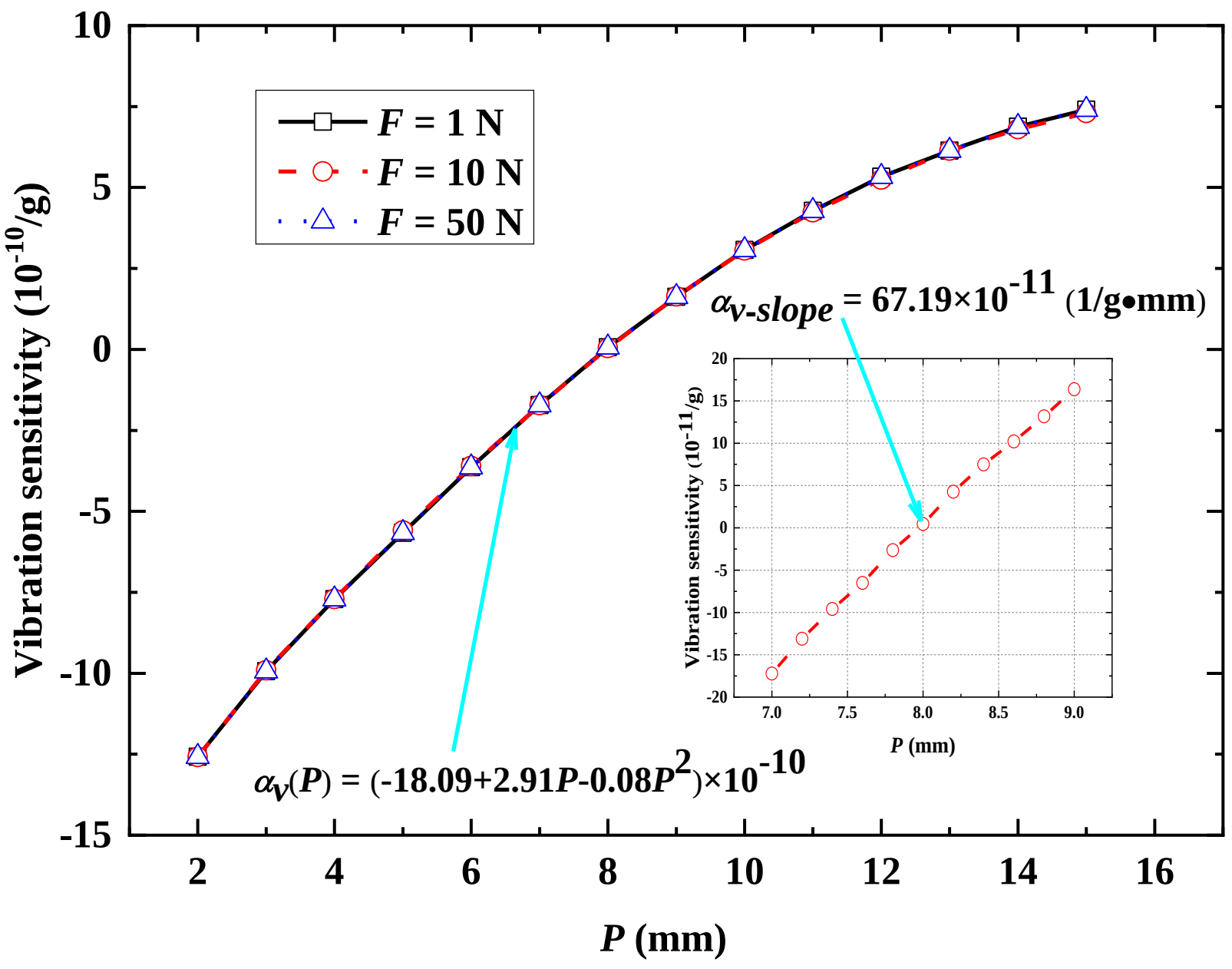

Fig.7 Simulation results for the vertical vibration sensitivity of the optical reference cavity as a function of the support point distance $P$ from mirror.

As illustrated in Fig.7, the vibration sensitivity of the optical reference cavity is plotted against the parameter $P$ under three different clamping force $F$. The three curves nearly overlap, demonstrating that within the clamping force $F$ range from 1 N to 50 N, the clamping force exerts negligible influence on the vibration sensitivity of the optical reference cavity. The overall vibration sensitivity variation follows a quadratic fitting function, showing a strong dependence on parameter $P$. Vibration sensitivity rises monotonically as $P$ increases and crosses zero at a certain critical distance. The magnified sub-plot further gives the local slope near the zero-crossing point. This nonzero local gradient implies that exact zero-vibration-sensitivity cannot be perfectly achieved in practical fabrication and assembly. Small machining or mounting deviations of distance $P$ will introduce non-negligible residual vibration sensitivity. Therefore, precise control over dimension $P$ is essential for realizing low-vibration-sensitive optical reference cavities.

### 3.3. Automatic frequency locking

Fig. 8 illustrates the automatic frequency-locking procedure developed for this laser. The sequence starts by initializing the laser PZT drive voltage and setting a threshold for the cavity transmission peak. The PZT voltage is then scanned continuously while the system monitors whether the transmission peak reaches the threshold. Once the threshold is met, OEO locking is established and the PI servo controller closes the frequency-stabilization loop. During locked operation the transmission peak is monitored in real time: if its amplitude falls below the threshold, the PZT scan restarts automatically while the servo controller remains ready to reclose the loop, restoring stabilization without user intervention.

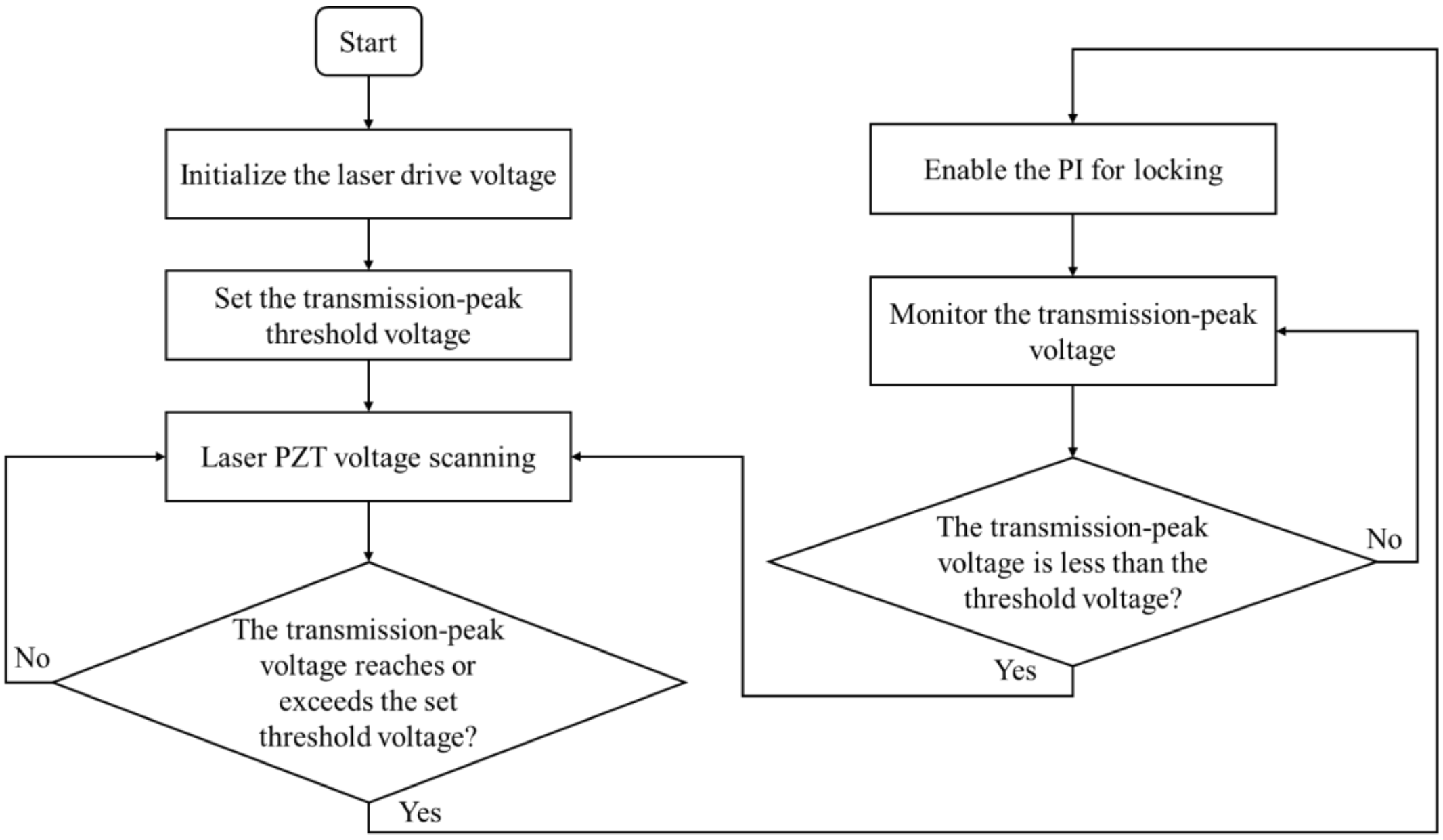

Fig.8 Procedure for automatic frequency locking of the ultra-stable laser.

To verify the repeatability of the automatic frequency-locking procedure, the ultra-stable laser is re-locked approximately 100 times, with the results presented in Fig. 9. Fig. 9 (b) shows a histogram of the re-locking time distribution, in which the median is approximately 1.14 s. Fig. 9 (a) presents the corresponding cumulative histogram, indicating that about 98% of the re-locking events fall within 1.5 s. In practical operation, occasional vibrations and other disturbances may induce abrupt amplitude jumps of the cavity transmission peak, which can mislead the monitoring logic and trigger undesired re-locking. To address this issue, the monitoring system implements a delayed-judgment strategy with a hold-off time of approximately 0.5 s.

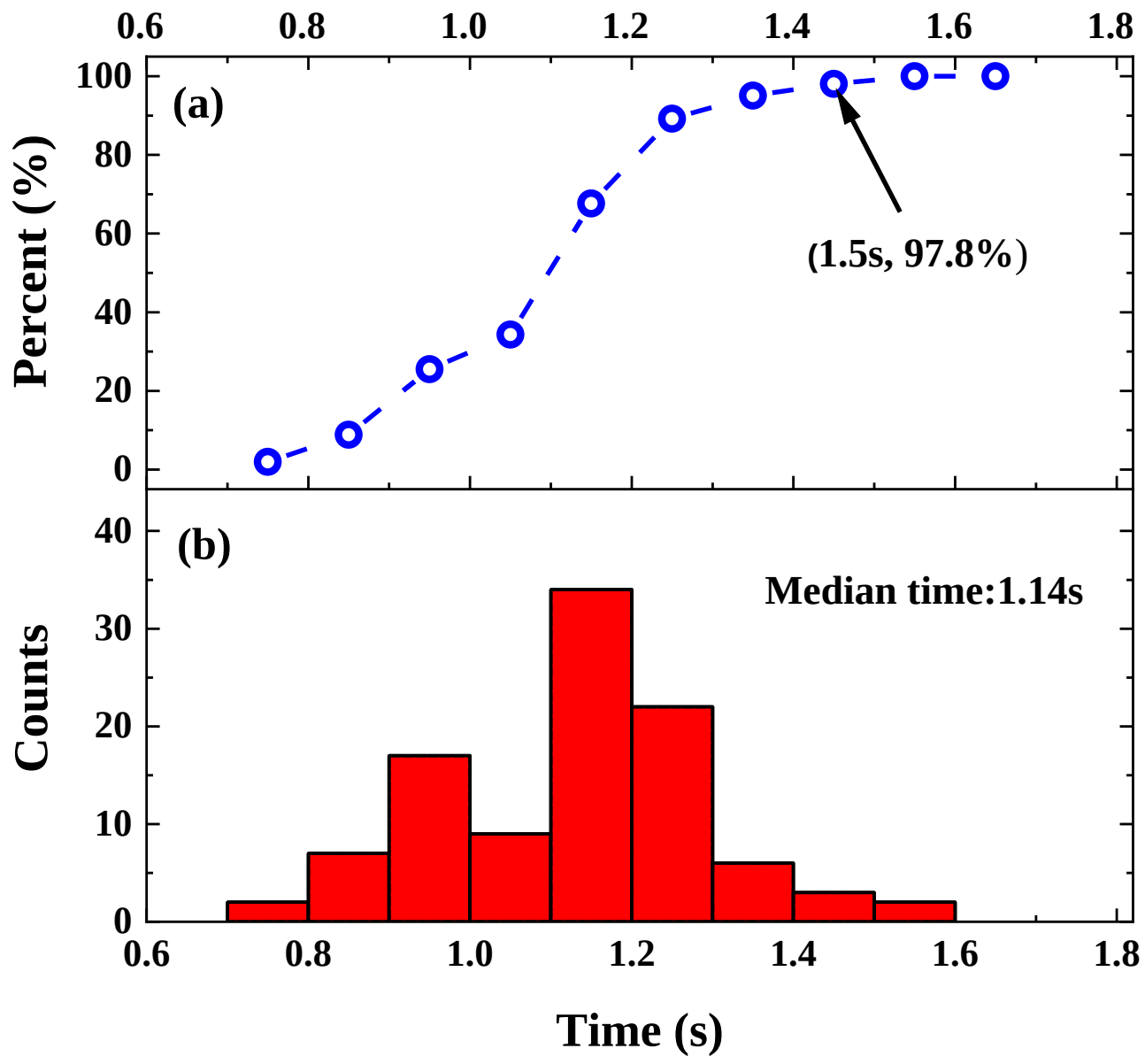

Fig. 9 Statistics of the automatic re-locking time over approximately 100 repeated locking trials. (a) Probability distributions. (b) Histogram of auto-locking time.

### 3.4. System integration

A photograph of the fully assembled system is shown in Fig. 10. All subsystems are integrated into a standard 2U rack-mount chassis with overall dimensions of approximately 600 mm × 482.6 mm × 88.9 mm and a total mass of ~10 kg. The front panel carries the FC/APC output port of the ultra-stable laser, a power switch, a control knob, and a touchscreen human-machine interface. The touchscreen displays the vacuum pressure and the frequency-locking status in real time, and allows adjustment of the PZT and EOM drive parameters, the PI servo controller settings, and the ion pump supply voltage. A 220 V AC mains inlet is located on the rear panel; during operation, the complete system draws approximately 10 W.

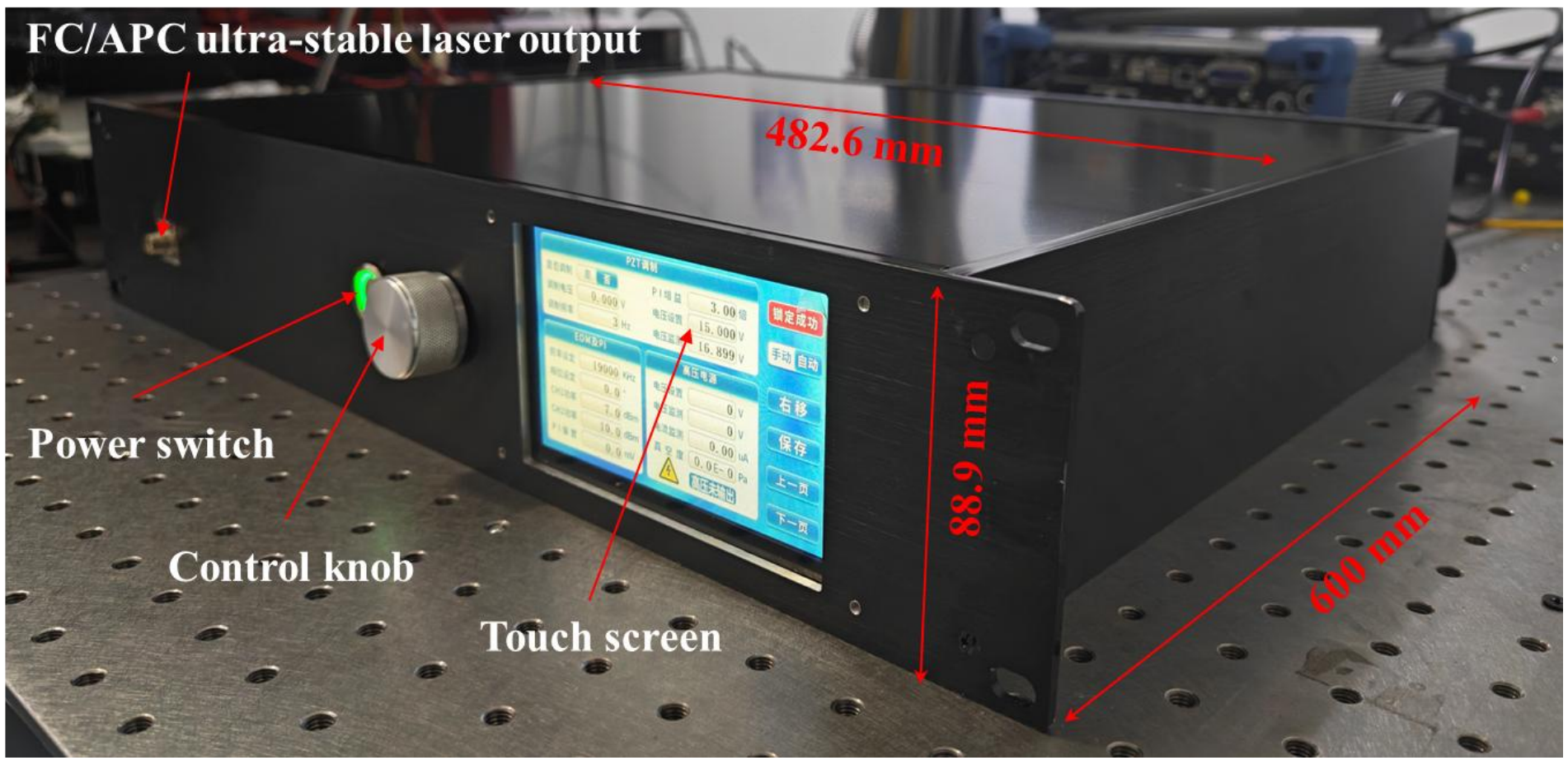

Fig.10 Photograph of the field-deployable ultra-stable laser system.

## 4. Results and Discussion

The performance of an ultra-stable laser is characterized by its phase noise, linewidth, and frequency stability. These parameters are evaluated through optical heterodyne beating between the developed laser and a second laboratory ultra-stable laser that exhibits a fractional frequency stability at the $10^{-16}$ level and a linewidth of ~0.26 Hz [35]. The outputs of the two lasers are combined with a fiber beam splitter, and the beat signal is detected by a high-speed photodetector. Given the short fiber lengths in this setup, no fiber-noise cancellation is applied.

### 4.1. Phase noise

Beat-note phase noise spectra are acquired with a Rohde & Schwarz FSUP26 signal and spectrum analyzer over Fourier frequencies from 0.1 Hz to 100 kHz and are reported as single-sideband phase noise in dBc/Hz. Fig. 11 plots four traces: the free-running laser (green solid line), the OEO-locked laser (red solid line), the theoretical OEO-locking noise from Eq. (8) (cyan short-dashed line), and the cavity thermal noise limit (black dashed line).

In free-running operation, the phase noise is high across the full measurement bandwidth,

reaching approximately 150 dBc/Hz at 0.1 Hz offset and ~100 dBc/Hz at 1 Hz. OEO locking suppresses the noise over the entire Fourier-frequency range: at 0.1 Hz the locked phase noise falls to ~30 dBc/Hz, a 120 dB reduction, and at 1 Hz it reaches ~0 dBc/Hz, a 100 dB improvement. From 1 to 100 Hz, the locked-laser phase noise closely approaches the cavity thermal noise floor, while the slight excess below 1 Hz is attributed to residual thermal perturbations. From 100 Hz to 100 kHz, the measured noise follows the theoretical OEO locking limit of Eq. (8); above several kilohertz it forms a plateau near −90 dBc/Hz, indicating effective suppression of technical excess noise. According to the cavity thermal-noise model and Eq. (8), further improvement can be pursued by adopting crystalline mirror coatings to lower the thermal-noise limit, increasing the cavity finesse, and shortening the physical length of the OEO loop.

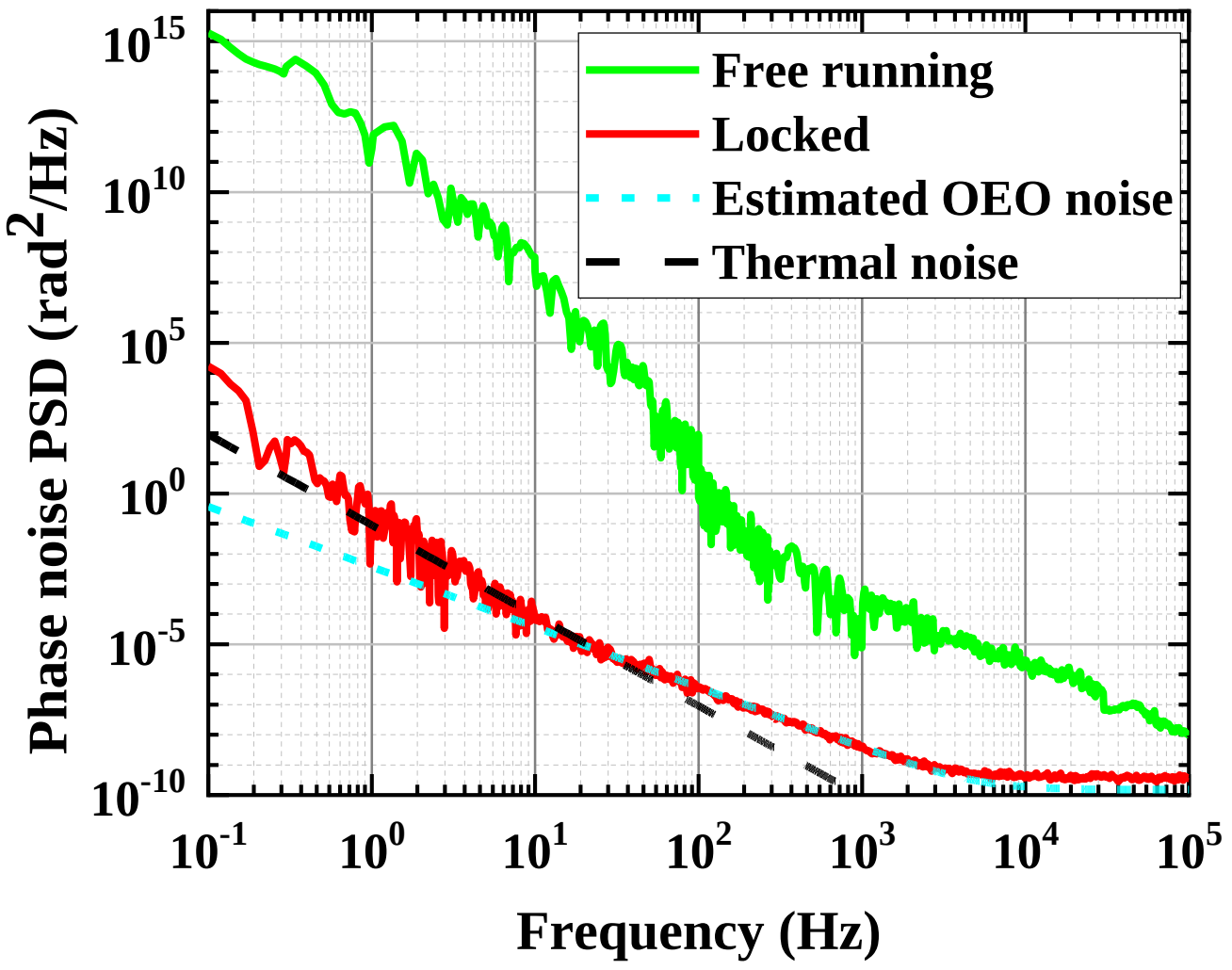

Fig.11 Measured phase noise for the ultra-stable laser.

### 4.2. Linewidth and frequency stability

The beat-note linewidth is characterized with an FFT-based dual-channel dynamic signal analyzer (Stanford Research Systems SR785). The beat signal is first down-converted to approximately 50 kHz with a frequency synthesizer and then acquired by the SR785. As shown in Fig. 12, a Lorentz fit yields a full width at half maximum (FWHM) linewidth of ~0.95 Hz, measured at a resolution bandwidth (RBW) of 0.3 Hz.

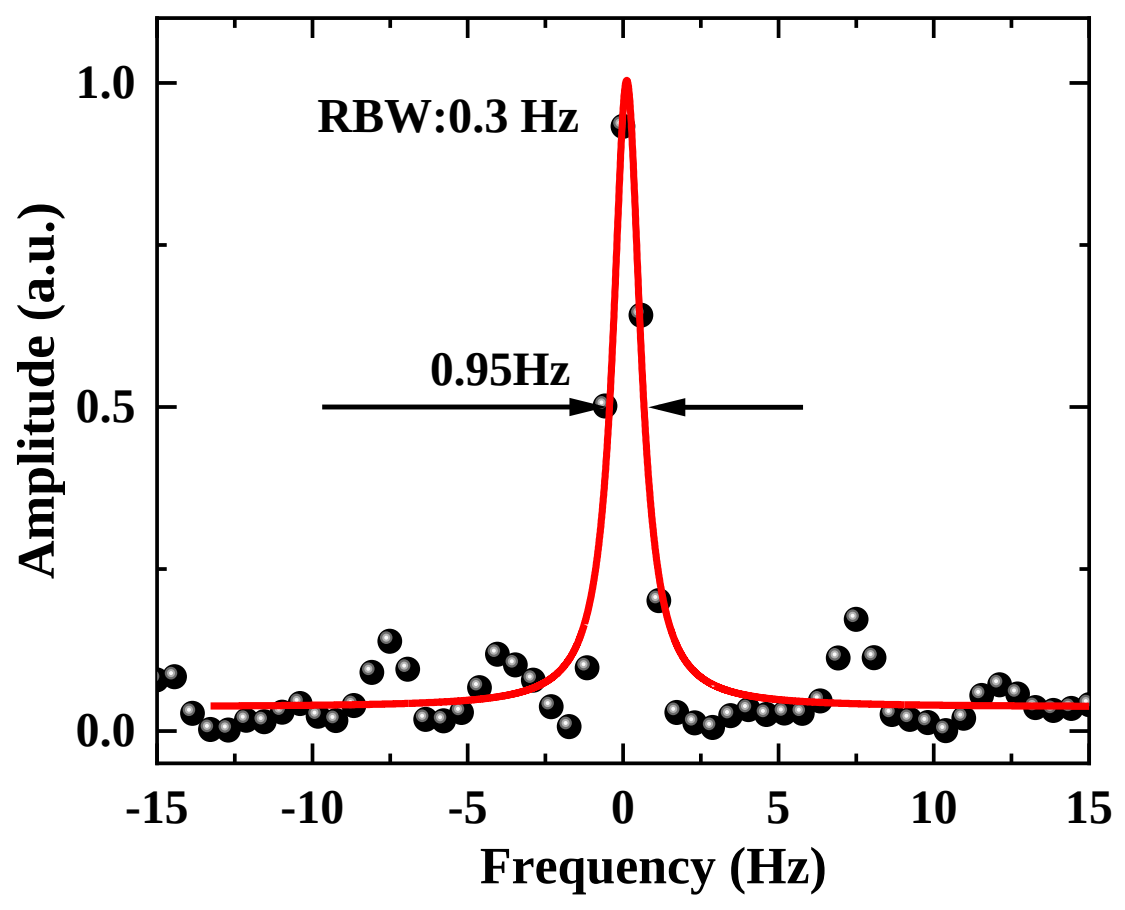

Fig.12 Beat note linewidth of the ultra-stable laser.

For ultra-stable lasers, F. L. Walls and A. E. Wainwright [36] propose that the linewidth of the laser can be derived from the phase change. When the accumulated phase noise of the laser reaches 1 rad², the linewidth of the laser is defined as $\Delta v$.

$$\int_{\frac{\Delta v}{2}}^{\infty} S_{\emptyset}(f)df = 1rad^2 \tag{11}$$

Integrating the measured phase noise spectral density (Fig. 10) yields the phase noise curve in Fig. 13, from which we compute an integrated linewidth of ~0.42 Hz. The beat note FWHM linewidth is roughly twice this integrated value. This discrepancy is mainly ascribed to residual slow laser frequency drift contributing to the direct beat note measurement.

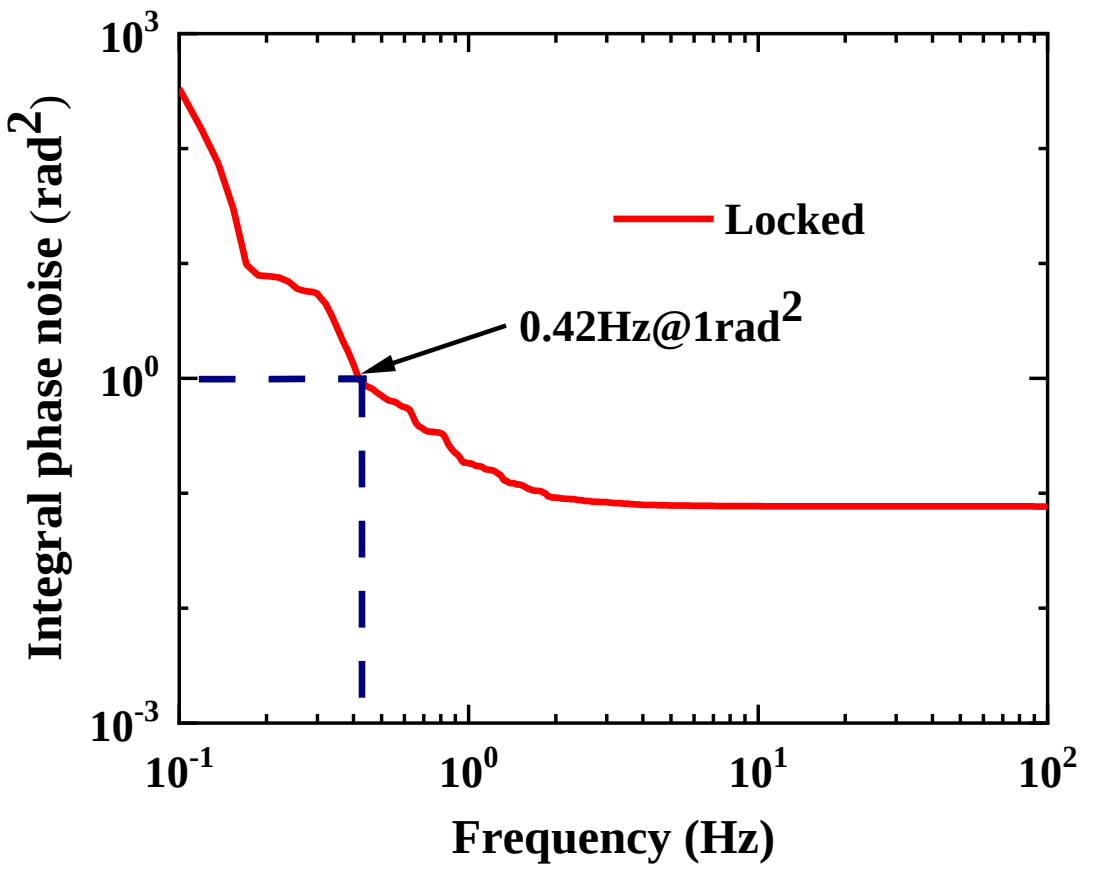

Fig.13 Integrated phase noise of the ultra-stable laser.

The fractional frequency stability is measured by recording the beat-note frequency fluctuations with a frequency counter (Keysight 53230A) at a 0.1 s gate time. The resulting modified Allan deviation (MADEV) is presented in Fig. 14. Between 0.2 s and 0.4 s, the measured MADEV approaches the cavity thermal-noise limit of ~$1.6\times10^{-15}$, reaching ~$2.2\times10^{-15}$ at 0.4 s; it is ~$5\times10^{-15}$ at 1 s, $1.5\times10^{-14}$ at 10 s, and $5.0\times10^{-14}$ at 100 s. The short-term stability is presently dominated by mechanical vibration; in future work, we plan to reduce the cavity vibration sensitivity toward the $10^{-11}$/g level to mitigate this contribution. The long-term stability is primarily limited by temperature-induced cavity drift. Planned upgrades include fabricating ULE cavities whose zero-crossing temperature lies above room temperature to relax the requirements of active temperature control, and adding a second thermal shield to further increase the thermal time constant. With these improvements, long-term fractional frequency stability at the $10^{-15}$ level is expected to be feasible.

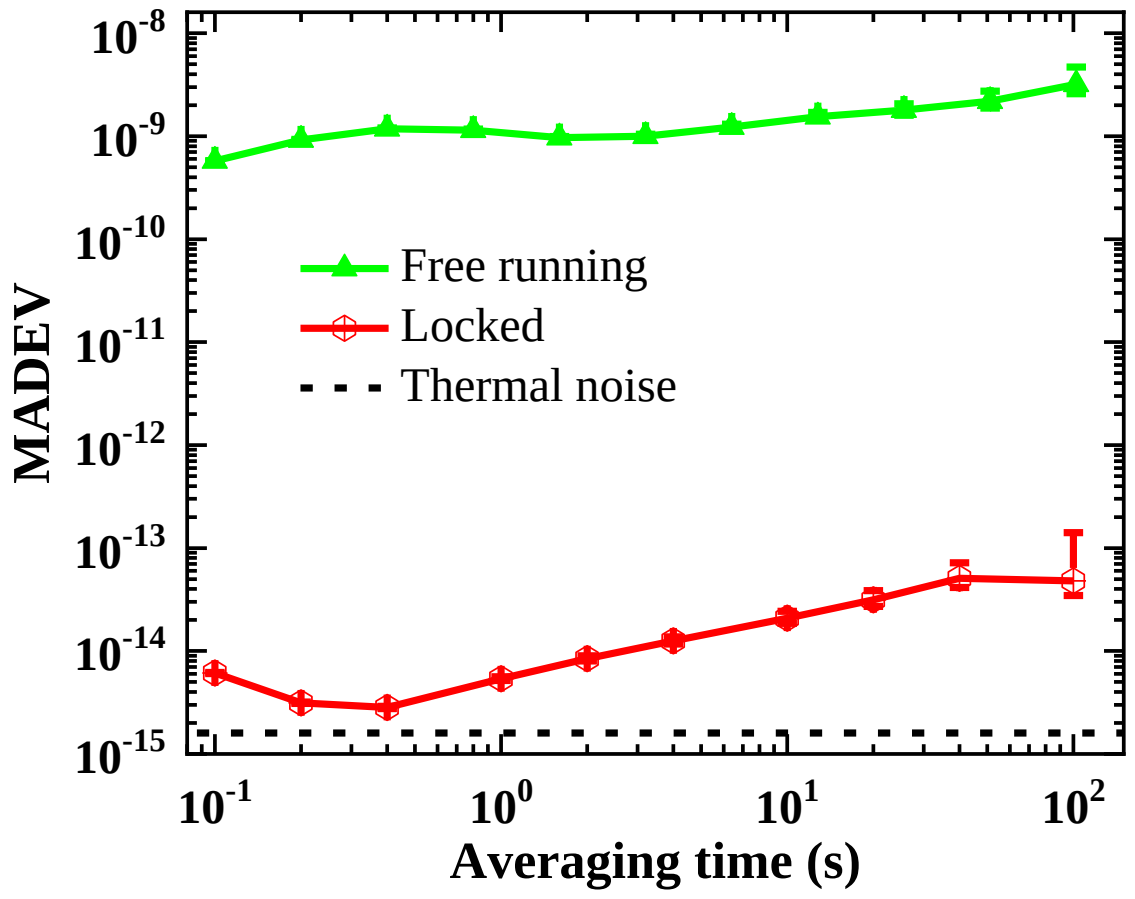

Fig.14 Fractional frequency stability of the ultra-stable laser. Red solid line: MADEV of the ultra-stable laser; black dashed line: thermal noise of the optical reference cavity; green solid line: MADEV of the free-running laser.

### 4.3. Performance comparison

The performance of representative compact ultra-stable laser systems reported in recent years is summarized in Table 3 and compared in Fig. 15 in terms of fractional frequency stability, linewidth, locking scheme, resonator volume, and system form factor. In terms of short-term stability, the proposed system achieves ~$5\times10^{-15}$ at 1 s averaging time. This result outperforms previously reported small-volume resonators, including the 0.7 mL whispering-gallery-mode (WGM) resonator ($5.2\times10^{-13}$ at 1 s) [44], the 4.5 mL monolithic Fabry–Pérot (FP) cavity ($9.1\times10^{-14}$) [39], and the 24 mL vacuum-gap FP cavity ($5.3\times10^{-14}$) [41], and is of the same order of magnitude as 125 mL vacuum-gap FP resonators (~$1.3$–$2.6\times10^{-15}$) [26,38,42], while the resonator volume is reduced to 19 mL. At longer averaging times, the measured stability of $1.5\times10^{-14}$ at 10 s and $5.0\times10^{-14}$ at 100 s remains comparable to that of the bulkier counterparts. In terms of linewidth, the integrated linewidth of ~0.42 Hz is narrower than the ~4 Hz of the 0.7 mL WGM resonator [44], the ~15 Hz of the monolithic FP cavity [39], and the ~3 Hz of the 24 mL vacuum-gap FP cavity [41]. Larger 125 mL cavities reach ~0.23–0.57 Hz [26,38], but those systems occupy up to 8U rack packaging, which limits field deployment. In terms of system integration, most previously reported $10^{-15}$-level systems rely on 125 mL resonators in 8U packages, whereas the instrument reported here integrates a 19 mL resonator and all control electronics within a single 2U chassis.

Overall, previously reported compact ultra-stable lasers have either reduced the cavity volume at the cost of degraded frequency stability, or retained $10^{-15}$-level stability with bulky, heavy 8U hardware. The present work achieves $10^{-15}$-level fractional frequency stability and a sub-hertz integrated linewidth with a 19 mL reference resonator, packaged in a 2U instrument, providing a practical route toward field-deployable precision metrology.

**Table 3 Performance comparison of compact ultra-stable laser systems reported in recent years.**

| Reference Resonator | Frequency Stability | Linewidth | Locking Method | Resonator Volume | System Form Factor | Year |
|---|---|---|---|---|---|---|
| **Vacuum-gap FP [37]** | $7\times10^{-15}$@1s<br>$1.1\times10^{-14}$@10s | / | PDH | 8mL | / | 2022 |
| **Vacuum-gap FP [38]** | $2.6\times10^{-15}$@1s<br>$3.1\times10^{-15}$@10s<br>$2.0\times10^{-14}$@100s | ~0.57 Hz | PDH | 125 mL | 8U | 2023 |
| **Monolithic FP [39]** | $9.1\times10^{-14}$@1s<br>$3.0\times10^{-13}$@10s | ~15 Hz | PDH | 4.5mL | / | 2024 |
| **Vacuum-gap FP [40]** | $2.4\times10^{-14}$@1s<br>$2\times10^{-13}$@10s<br>$1.5\times10^{-12}$@100s | / | PDH | 9.7mL | / | 2024 |
| **Vacuum-gap FP [26]** | $1.8\times10^{-15}$@1s<br>$2.0\times10^{-15}$@10s<br>$3.0\times10^{-15}$@100s | ~0.23 Hz (Integral) | OEO | 125 mL | / | 2024 |
| **Vacuum-gap FP [41]** | $5.3\times10^{-14}$@1s<br>$5.6\times10^{-14}$@10s<br>$9.0\times10^{-14}$@100s | ~3 Hz | PDH | 24mL | 4U | / |
| **Vacuum-gap FP [42]** | $1.3\times10^{-15}$@1s<br>$2.5\times10^{-15}$@10s<br>$9.9\times10^{-15}$@100s | ~2 Hz | PDH | 125mL | 8U | / |
| **Vacuum-gap FP [43]** | $5.0\times10^{-12}$@1s<br>$1.4\times10^{-12}$@10s | ~150 Hz (Integral) | Self-injection | 0.6 mL | / | 2025 |
| **WGM [44]** | $5.2\times10^{-13}$@1s<br>$4.0\times10^{-12}$@10s<br>$3.1\times10^{-11}$@100s | ~4 Hz (Integral) | PDH | 0.7mL | / | 2026 |
| **This work** | $5\times10^{-15}$@1s<br>$1.5\times10^{-14}$@10s<br>$5.0\times10^{-14}$@100s | ~0.42 Hz (Integral) | OEO | 19 mL | 2U | 2026 |

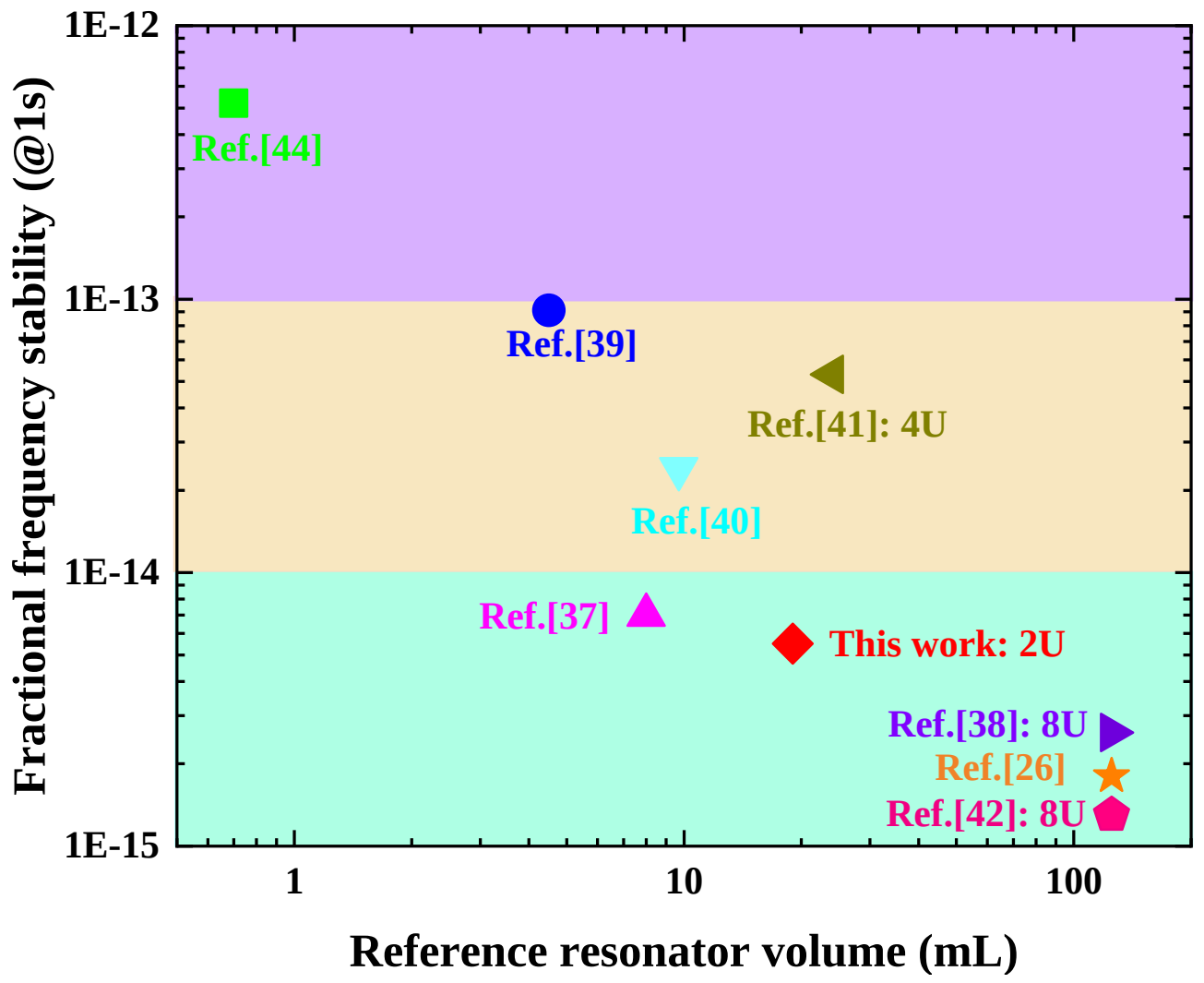

Fig.15 Comparison of fractional frequency stability @1 s versus reference resonator volume for compact ultra-stable lasers.

## 5. Conclusions

We have demonstrated, to our knowledge, the most compact field-deployable ultra-stable laser based on optoelectronic oscillator (OEO) locking, integrated into a standard 2U rack chassis. A 50 mm-long, 19 mL optical reference cavity with low vibration sensitivity and the thermal shield stabilizes a 1550 nm fiber laser with ~1 kHz free-running linewidth to the cavity thermal-noise limit. A hybrid analog-digital controller realizes fully automatic frequency locking and autonomous re-locking for unattended operation. The laser delivers an integrated linewidth of ~0.42 Hz and a fractional frequency stability of ~$5\times10^{-15}$ at 1 s averaging time (~$2.2\times10^{-15}$ at 0.4 s). Measured phase noise approaches the cavity thermal noise limit below 100 Hz and follows the theoretical OEO locking limit from 100 Hz to 100 kHz. The overall system volume is approximately one quarter that of previously reported integrated, portable ultra-stable lasers of comparable performance. Future work will target lower cavity vibration sensitivity, crystalline mirror coatings, higher finesse, a shorter OEO loop, and improved thermal shield, with the aim of reaching $10^{-15}$-level long-term frequency stability. These advances make the present laser a practical tool for field-deployable precision metrology, including optical clocks, coherent sensing, and quantum technologies.

**Funding.** This project was supported by the National Natural Science Foundation of China (Grant No. 12303076 and 12403078), the Science and Technology Innovation Team of Shaanxi Province (Grant No. 2026RS-CXTD-01), the Key Research and Development Program of Shaanxi Province (Grant No. 2025CY-YBXM-0060), the Sanqin Talents' Special Support Program (Grant No. 09R0557A00), the Shaanxi Provincial Outstanding Young Scientist Fund Project (Grant No. 2025JC-JCQN-082), the Shaanxi Province Qin Chuangyuan "Scientist + Engineer" Team Construction Project (Grant No. 2025QCY-KXJ-162 and 2026PT-KXJ-003).

**Acknowledgments.** The authors would like to thank Fuyu Sun from the National Time Service Center, Chinese Academy of Sciences, for valuable discussions.

**Disclosures.** The authors declare no conflicts of interest.

**Data availability.** Data underlying the results presented in this study are not publicly available at this time but may be obtained from the authors upon a reasonable request.